\documentclass[twocolumn,amsmath,amssymb,10pt,superscriptaddress,a4paper,letterpaper,fleqn]{revtex4-1}
\usepackage{amssymb}
\usepackage{epsfig}
\usepackage{graphicx}
\usepackage{dcolumn}
\usepackage{array}
\usepackage{bm}
\usepackage{fancyheadings}
\usepackage{longtable}
\usepackage{multirow}
\usepackage{float}
\usepackage{afterpage}
\usepackage{color}

\setlongtables
\usepackage[breaklinks=true,linkbordercolor={1 1 1}]{hyperref}

\begin{document}
\setcounter{page}{1}


\title{
\qquad \\ \qquad \\ \qquad \\  \qquad \\  \qquad \\ \qquad \\
Propagation of nuclear data uncertainties for ELECTRA burn-up calculations}

\author{Henrik Sj\"ostrand}
\email[Corresponding author: ]{henrik.sjostrand@physics.uu.se}
\affiliation{Division of Applied Nuclear Physics, Department of Physics and Astronomy, Uppsala University, Uppsala, Sweden}

\author{Erwin Alhassan}
\affiliation{Division of Applied Nuclear Physics, Department of Physics and Astronomy, Uppsala University, Uppsala, Sweden}

\author{Junfeng Duan}
\affiliation{Division of Applied Nuclear Physics, Department of Physics and Astronomy, Uppsala University, Uppsala, Sweden}

\author{Cecilia Gustavsson}
\affiliation{Division of Applied Nuclear Physics, Department of Physics and Astronomy, Uppsala University, Uppsala, Sweden}

\author{Arjan Koning}
\affiliation{Division of Applied Nuclear Physics, Department of Physics and Astronomy, Uppsala University, Uppsala, Sweden}
\affiliation{Nuclear Research and Consultancy Group (NRG), Petten, The Netherlands}

\author{Stephan Pomp}
\affiliation{Division of Applied Nuclear Physics, Department of Physics and Astronomy, Uppsala University, Uppsala, Sweden}

\author{Dimitri Rochman}
\affiliation{Nuclear Research and Consultancy Group (NRG), Petten, The Netherlands}

\author{Michael \"Osterlund}
\affiliation{Division of Applied Nuclear Physics, Department of Physics and Astronomy, Uppsala University, Uppsala, Sweden}

\date{\today}

\begin{abstract}
{The European Lead-Cooled Training Reactor (ELECTRA) has been proposed as a training reactor for fast systems within the Swedish nuclear program. It is a low-power fast reactor cooled by pure liquid lead. In this work, we propagate the uncertainties in $^{239}Pu$ transport data to uncertainties in the fuel inventory of ELECTRA during the reactor life using the Total Monte Carlo approach (TMC). Within the TENDL project the nuclear models input parameters were randomized within their uncertainties and 740 $^{239}Pu$ nuclear data libraries were generated. These libraries are used as inputs to reactor codes, in our case SERPENT, to perform uncertainty analysis of nuclear reactor inventory during burn-up. The uncertainty in the inventory determines uncertainties in:  the long-term radio-toxicity, the decay heat, the evolution of reactivity parameters, gas pressure and volatile fission product content. In this work, a methodology called fast TMC is utilized, which reduces the overall calculation time. The uncertainty in the long-term radiotoxicity, decay heat, gas pressure  and volatile fission products were found to be insignificant. However, the uncertainty of some minor actinides were observed to be rather large and therefore their impact on multiple recycling should be investigated further.  It was also found that, criticality benchmarks can be used to reduce inventory uncertainties due to nuclear data. Further studies are needed to include fission yield uncertainties, more isotopes, and a larger set of benchmarks.}
\end{abstract}
\maketitle

\rhead{Henrik Sj\"ostrand \textit{et al.}}
\lfoot{}
\rfoot{}
\renewcommand{\footrulewidth}{0.4pt}


\section{Introduction}
Accurate and reliable nuclear data (ND) and their uncertainties are important in the design, modeling and development of GEN-IV reactors.  Out of the six nuclear reactors selected by the GEN-IV International Forum, three are fast spectrum reactors which have an added advantage for transmutation of minor actinides such as, americium and curium\cite{01GIF}.  The transmutation performances are significantly affected by nuclear data and it is important to quantify the resulting uncertainties on the integral parameters on fast reactor systems since their fuel can be heavily loaded with minor actinides\cite{01Paul}. During reactor operation, many nuclides are produced as a result of different types of neutron-nuclear interactions. The long-term changes in the properties of a nuclear reactor core over its lifetime can be determined by these changes in fuel composition due to fuel burnup. Knowledge of isotopic composition during reactor life time is important for the estimation of long-term radiotoxicity and decay heat of discharged fuel, the evolution of reactivity parameters, build up of gas pressure, and inventory of volatile fission product content (which constitute the source term in accident scenarios). For transport and storage of spent fuel both an accurate prediction of the total decay heat is essential, as well as the uncertainty in $k_{eff}$. These are the motivations to investigate the impact of nuclear data uncertainties in burnup for the European Lead Cooled Training reactor (ELECTRA)\cite{06wal}, which is a low power (0.5 MW) reactor intended for education and training purposes. To allow for a small core with cooling accomplished by natural circulation an inert matrix zirconium, plutonium nitride fuel has been chosen for ELECTRA. The fuel is composed of 60 mol\% ZrN and 40 mol\% of PuN with a TRU vector ($^{238-242}Pu$, $^{241}Am$: 3.5/ 51.9 / 23.8 / 11.7 / 7.9 / 1.2 [at. \%]), which resembles typical spent fuel available at the Swedish interim storage \cite{06wal}. The Total Monte Carlo method  \cite{08koning} was applied to study the effects of uncertainties in $^{239}Pu$ transport and activation data on the ELECTRA inventory during burn-up. In order to study the effects of transport nuclear data uncertainties 740 $^{239}Pu$ files from the TENDL project\cite{06Dim} have been used.  These files include information on transport data, i.e. data necessary for the calculating neutron transport in steady state (cross-sections, angular distributions, differential data, emission spectra and nu-bar). The same cross-section data was used as activation data in the depletion calculations.

\section{Total Monte Carlo approach}
In the Total Monte Carlo approach \cite{08koning} uncertainties in nuclear model parameter are derived from theoretical considerations. They are randomly sampled multiple times and each set of parameters are used in a suit of nuclear data codes: TALYS, TANES, TARES, TAFIS \cite{06Dim}. The output is a large set of random nuclear data files which are processed into ENDF format using the TEFAL code\cite{08koning}. These can then be compared with experimental data and some are discarded if the deviation from experimental data is too large. The accepted random files are processed using the ACER module in NJOY, which converts the random ENDF nuclear data files into ACE files. Theses random files can then be used in neutron transport and activation codes such as SERPENT \cite{06lepp} or MCNP.  The variation of the nuclear data causes distributions in the obtained quantities such as $k_{eff}$, inventory, temperature feedback coefficients, and kinetic parameters etc. \cite{08koning}.  The spread in these distributions ($\sigma_{obs}$) is both due to variations in ND in the different random files and due to the statistical uncertainties from the Monte Carlo transport calculation:
\begin{equation}
\sigma^2 _{obs}=\sigma^2_{ND}+\sigma^2_{statistics}
\label{sig}
\end{equation}
	 	
Since the transport code calculates $\sigma_{stat}$, it is possible to infer $\sigma_{ND}$, which is an estimated of the uncertainty due to ND in the investigated quantity.

\section{Simulations}
Burnup calculations were performed using the integrated neutron transport and depletion code SERPENT 1.1.17 \cite{06lepp}. One advantage of using SERPENT is that no distinction is made between the transport nuclear data used for calculating e.g. the flux spectrum and the activation nuclear data (cross-sections) used for solving the Bateman equation. Therefore the same random files are used both in the transport part and depletion part of the code. Consequently, full information on the correlations in nuclear data uncertainty propagation between these two steps is obtained. As mentioned earlier 740 $^{239}Pu$ files were obtained from the TENDL project and processed into ACE files with Njoy99.336 code.  JEFF3.1 general purpose library was used for all the other isotopes. The other input data used in this work are the decay and fission yield data read from the JEFF-3.1 nuclear data library. These files were not varied in this work. The calculations were performed for 20 inactive and 50 active cycles with 5000 neutrons/cycle. Burnup calculations were performed by invoking the predictor-corrector module in SERPENT for eight burnup steps from 0-55 Mwd/kg corresponding to the beginning of life (BOL) and end of life (EOL) of the reactor respectively\cite{06wal}. For decay calculations, eighteen steps were used from $0-10^{6}$ years after the reactor EOL. Continuous full power operation is assumed, up till the end of life (EOL) of the reactor. The EOL is reached when the reactivity drums has reached their end position. This was calculated to be at 55 at MWd/kg in ref. \cite{06wal}, which occurs after 7442 full power days. In the simulations performed for this paper the drums were not rotated but the simulation were executed with the drums in their start position.     
	 
\section{Fast TMC in burnup calculations}
\begin{figure}[htb]
\includegraphics[width=\columnwidth]{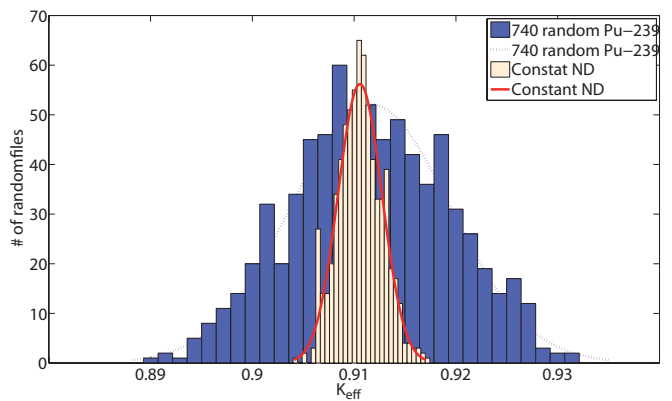}
\caption{The distribution in $k_{eff}$ in the two sets of simulations. The blue bars represent the 740 random files, where each simulation has unique $^{239}Pu$ ND. The light bars represent the outcome from the 700 simulations where the ND was kept constant. All simulations have unique seeds.}
\label{fig1}
\end{figure}

In ref. \cite{07Dim} fast TMC methodologies are described in detail and in this section only the burnup aspect will be discussed. From Eq. (1) we see that we need an estimate on $\sigma_{stat}$, but statistical uncertainties are not propagated in burnup codes. Therefore we run our transport code in two sets with the only difference that in the first set the code is run N1 times with constant ND and in the second set the code is run the N2 times, each time with a different random file (N2 = the number of random files). In our case N1= 700 and N2 = 740. FIG.~\ref{fig1} shows the result, where the set with constant nuclear data has a narrower distribution. Its spread is only due to statistics in the calculation, $\sigma_{stat}$. The set that correspond to the 740 random files  has a wider distribution and its spread, $\sigma_{obs}$, is both due to statistics and ND as outlined in Eq.(1). With this information $\sigma_{ND}$ can be inferred. Since we have many random files, the uncertainties on the spreads ($\Delta {\sigma}_{stat}$ and $\Delta {\sigma}_{obs}$) are small. As a consequence, we can reduce the statistical requirement on each run, lowering the number of neutrons, hence reducing the computation time. This is why the methodology is referred to as fast TMC. In this work the full core burnup calculation took 24 h on 60 cores. 

It is also of interest to know $\Delta {\sigma}_{ND}$, i.e., the uncertainty on the uncertainty due to nuclear data.  We first compute the  uncertainty of the variance due to nuclear data ($V_{ND}$): 

\begin{equation}
{\left( {\Delta {V_{ND}}} \right)^2} = {\left( {\Delta {V_{obs}}} \right)^2} + {\left( {\Delta V_{stat}^{}} \right)^2}
\label{Yi} 
\end{equation}

Where $V_{obs}$ is the variance in the random files, $V_{stat}$ is the variance for the set with constant nuclear data and $\Delta$ is the associated uncertainties.  The uncertainty in $V_{obs}$ is:     

\begin{equation}
\Delta {V_{obs}} = \sqrt 2 \frac{{S_{obs}^{}}}{{\sqrt {N_2^{}} }}
\label{delt}
\end{equation}

Where capital S is the sample variance. The uncertainty of  $V_{stat}$ is: 
\begin{equation}
\Delta V_{stat}^{} = \sqrt 2 \frac{{S_{stat}^{}}}{{\sqrt {N_1^{}} }}
\label{Y2}
\end{equation}

Combining the above equations: 
\begin{equation}
\left( {\Delta V_{ND}^{}} \right)_{}^2 = \left( {\sqrt 2 \frac{{S_{obs}^{}}}{{\sqrt {N_2^{}} }}} \right)_{}^2 + \left( {\sqrt 2 \frac{{S_{stat}^{}}}{{\sqrt {N_1^{}} }}} \right)_{}^2
\label{Yi}
\end{equation}
Normally, what is asked for is the uncertainty in uncertainty due to ND , which is: 

\begin{align}
\Delta \sigma _{ND}^{} = \frac{{\Delta V_{ND}^{}}}{{2\sigma _{ND}^{}}} \approx \frac{{\sqrt {\frac{{S_{obs}^2}}{{{N_2}}} + \frac{{S_{stat}^2}}{{{N_1}}}} }}{{\sqrt {2\left( {S_{obs} - S_{stat}} \right)} }} \nonumber\\
 = [\frac{{{\rm{S}}_{{\rm{obs}}}^{\rm{2}}}}{{{{\rm{N}}_2}}}{\rm{  >  >  }}\frac{{{\rm{S}}_{{\rm{stat}}}^{\rm{2}}}}{{{N_1}}}{\rm{ and }}S_{obs}^2{\rm{  >  >  }}S_{stat}^2{\rm{ }}]  \approx  \frac{{\sigma _{ND}^{}}}{{\sqrt {2N_2^{}} }}
\end{align}

As can be observed from the equation it is more important to have many random files than to improve the statistical spread in each run. The number of random files should normally be higher than the number of runs with constant ND.

\section{Results and Discussion}
\begin{figure}[htb]
\includegraphics[width=\columnwidth]{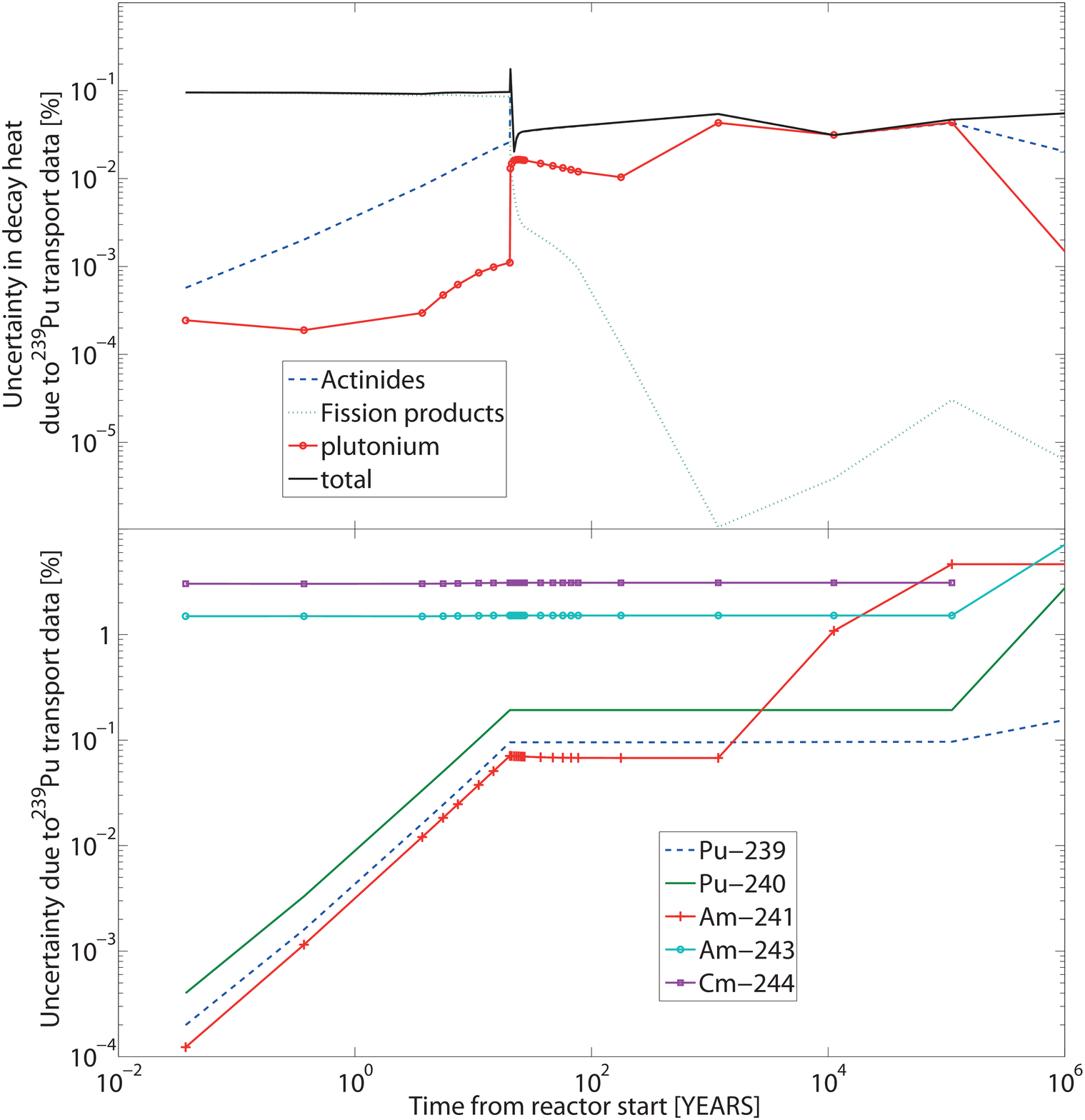}
\caption{(Top) relative decay heat uncertainties due to transport (and activation) $^{239}Pu$ data (\%) as a function of time (years). (Bottom)  relative uncertainties in the inventory due to transport (and activation) $^{239}Pu$ data for some selected isotopes as a function of time (years). At 20 years the reactor reaches its EOL.}
\label{fig2}
\end{figure}

In previous work on ELECTRA, \cite{06wal} it was observed that the total reactivity loss at EOL was $9800\pm200$ pcm  with 13.8\% of the initial plutonium converted into fission products or americium. In FIG.~\ref{fig2}, we present the main results on the uncertainty in inventory and uncertainty in decay heat.  When we look at the evolution during burnup it is seen that there is only small uncertainties for major actinides and consequently there is only a small uncertainty in the radiotoxicity. For the minor actinides larger uncertainties are observed e.g.  11\% in $^{249}Cf$.  This has little effect on integral parameters such as radiotoxicity (not shown) or decay heat (FIG.~\ref{fig2}, top). However the large uncertainties in the minor actinides due to $^{239}Pu$ data should be investigated in more detail since it might have effects in multiple recycling scenarios, both on decay heat and radiotoxicity as well as on different safety parameters such as the Doppler coefficient.  

\subsection{Benchmarks for inventory}
\begin{figure}[htb]
\includegraphics[width=\columnwidth]{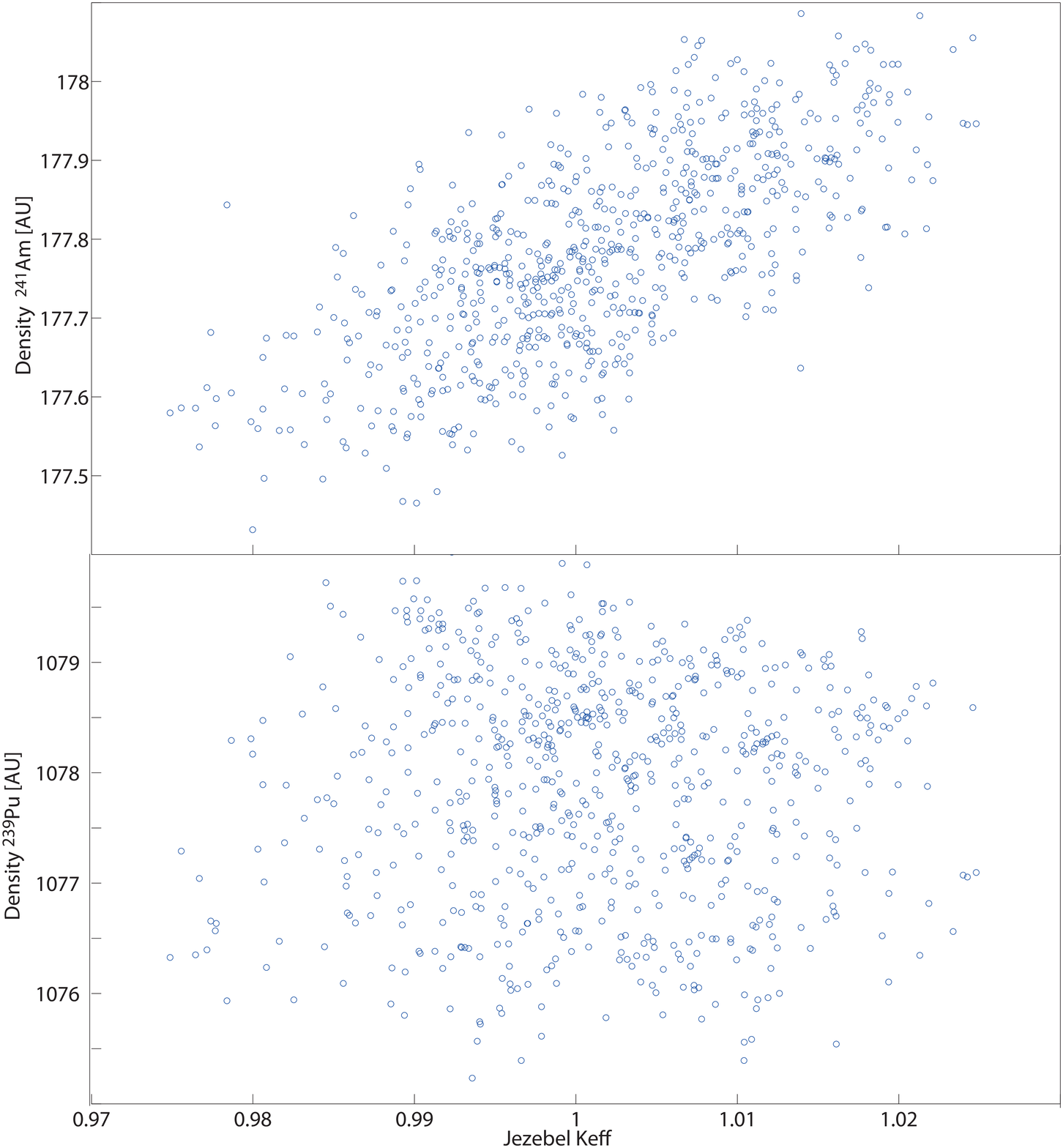}
\caption{The $^{241}Am$ (top) and $^{239}Pu$ (bottom) inventories for the 740 $^{239}Pu$ random files are plotted (each blue circle representing one random file) against the calculated $k_{eff}$ in the Jezebel Benchmark for the same files. The Jezebel benchmark has a nominal value of 1.000.}
\label{fig3}

\includegraphics[width=\columnwidth]{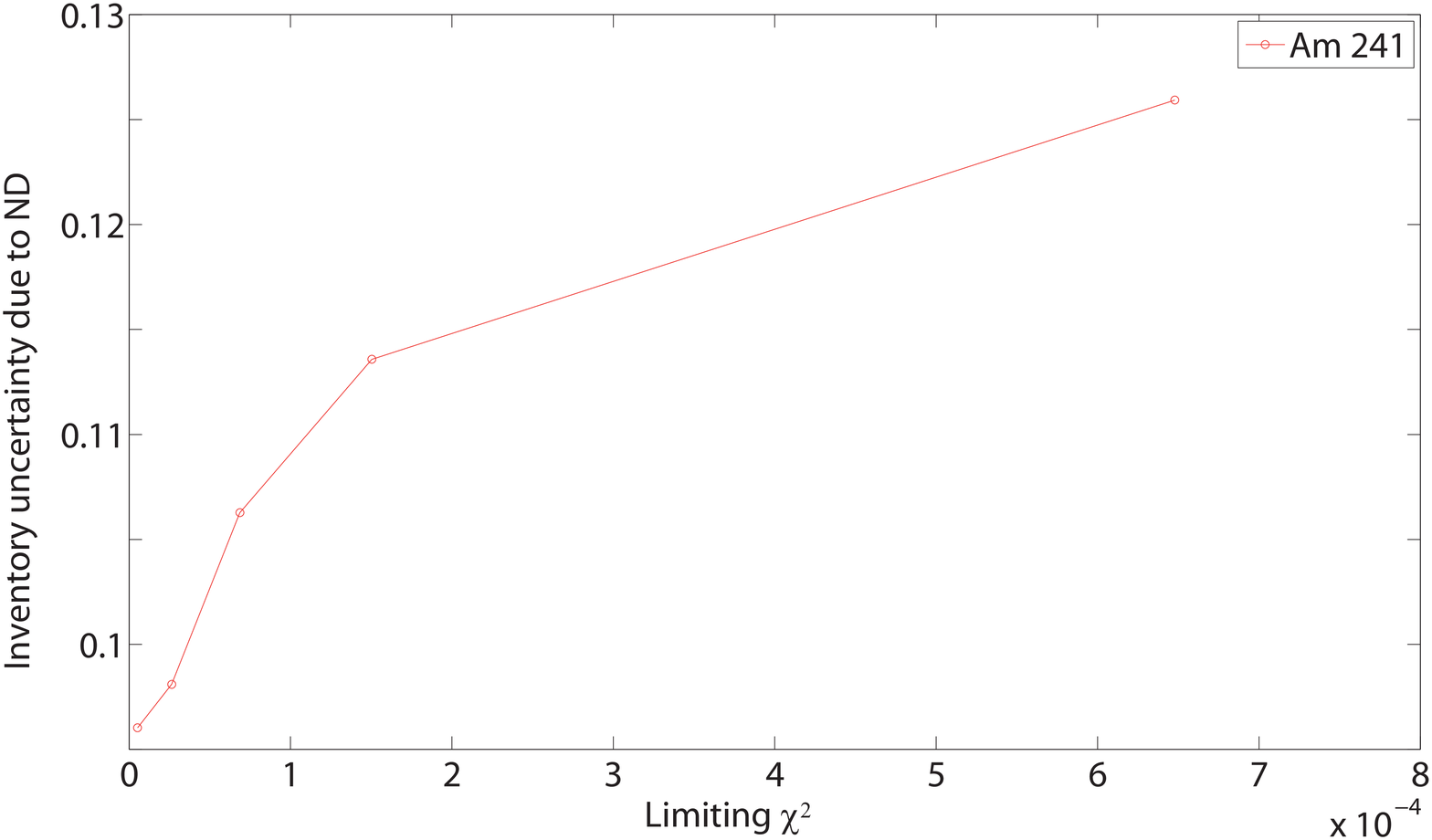}
\caption{The uncertainty in $^{241}Am$ inventory due to nuclear data vs. the maximum allowed $\chi^2$ for the random file on the Jezebel benchmark.}
\label{fig4}
\end{figure}

In ref. \cite{06Dim} benchmarks are used to select the best random file and in ref. \cite{06Al}, a new method to reduce the uncertainty in $k_{eff}$ due to nuclear data using criticality benchmarks is presented. In this work we use the same methodology to investigate if criticality benchmarks can be used to reduce uncertainties in the inventories for the different isotopes. The Jezebel benchmark \cite{06Bri} was used as the criticality benchmark. All random files were also run on that benchmark.  To use the benchmark to reduce the uncertainty, a correlation between the benchmark and the investigated parameter (in this case the number density) has to be found. FIG.~\ref{fig3} shows examples of correlations between the number density in $^{241}Am$ and $^{239}Pu$ and the $k_{eff}$ in Jezebel for the different benchmarks.  A correlation exists between the $^{241}Am$ number density and Jezebel $k_{eff}$. Consequently, the Jezebel Benchmark can be used to reduce uncertainties in the $^{241}Am$ inventory. This is not the case for the $^{239}Pu$ inventory. 

For each random file a goodness number is calculated $\chi^2$: 
\begin{equation}
\chi^{2}_{j}=\frac{(C_{j}-E)^{2}}{C_{j}}
\label{chi} 
\end{equation}

Where $C_{j}$ is the calculated $k_{eff}$ for the the random file $j$, and $E$ is the nominal value for the benchmark, in this case 1.0000. An acceptance criteria based on $\chi^2$ can be used, which is illustrated in FIG.~\ref{fig4}, where it is shown that with a more rigid acceptance criteria the uncertainty due to ND is reduced. It is found that a 25\% reduction in uncertainty is achieved for $^{241}Am$, however no significant reduction in uncertainty was achieved for $^{239}Pu$. All actinides were tested for correlations in all burnup steps and in most cases, except for $^{239}Pu$ and $^{240}Pu$, correlations were observed.  It was found that the correlation was quite constant with burnup.

\section{Conclusions and outlook}
Using SERPENT, fast TMC has successfully been performed for burnup in a full core 3-D model of ELECTRA using 740 $^{239}Pu$ transport-data random files. This was shown to be achievable with quite limited computer resources, showing the strength of fast TMC.  In the work the same random files were used in the depletion calculation and the transport calculation. It was found that the uncertainty in the long-term radiotoxicity, decay heat, gas pressure and volatile fission products is small for $^{239}Pu$ transport data uncertainties. The uncertainties on some minor actinides are large enough that their impact on safety parameters in multiple recycling should be investigated further. Decay data and fission yield data was not varied. A next step would be to include fission yield uncertainties by e.g. combining  TALYS and GEF codes. 
Criticality benchmarks can be used to select acceptable ND files and thereby reduce inventory uncertainties. In a next step a larger set of benchmarks should be used.  It was found that the Jezebel benchmark could be used to reduce the uncertainty in inventory for some isotopes, but not for all. The reason for this has to be investigated further. 

\end{document}